# Effect of crystallinity on spin-orbit torque in 5d iridium oxide IrO$_2$

Tetsuro Morimoto[1], Kohei Ueda[1,2,3,a)], Junichi Shiogai[1,3], Takanori Kida[4], Masayuki Hagiwara[4], and Jobu Matsuno[1,2,3]

[1]Department of Physics, Graduate School of Science, Osaka University, Osaka 560-0043, Japan

[2]Center for Spintronics Research Network, Graduate School of Engineering Science, Osaka University, Osaka 560-8531, Japan

[3]Spintronics Research Network Division, Institute for Open and Transdisciplinary Research Initiatives, Osaka University, Osaka, 565-0871, Japan

[4]Center for Advanced High Magnetic Field Science, Graduate School of Science, Osaka University, Osaka 560-0043, Japan

a)Author to whom correspondence should be addressed: kueda@phys.sci.osaka-u.ac.jp

**ABSTRACT**

The 5d transition-metal oxides provide an intriguing platform for generating an efficient spin current due to a unique electronic structure dominated by 5d electrons with strong spin-orbit coupling. Here, we report on the effect of crystallinity on current-driven spin-orbit torque (SOT) in binary 5d iridium oxide IrO$_2$ thin films by controlling amorphous, polycrystalline, and epitaxial states. By conducting harmonic Hall measurement in bilayers composed of ferromagnetic Co$_{20}$Fe$_{60}$B$_{20}$ and IrO$_2$, we find that dampinglike (DL) SOT is larger than fieldlike SOT for all the samples. We also demonstrate that both electrical resistivity and the DL SOT efficiency increase in order of epitaxial, polycrystalline, and amorphous IrO$_2$. Despite their different electrical conductivities, spin Hall conductivities of the three states of the IrO$_2$ layer are found to be nearly constant, which is consistent with the intrinsic regime of the spin Hall effect scaling relation. Our results highlight the important role that crystallinity plays in the spin-current generation, leading to the potential technological development of spintronic devices based on the 5d transition-metal oxides. (169 words)

Spin-orbit coupling (SOC) is a relativistic effect that couples the orbital angular momentum of an electron with its spin angular momentum. Interfaces in bilayers composed of ferromagnet (FM) and non-magnet (NM) with a strong SOC provide a rich playground for exploring emergent spin related phenomena and the underlying physics [1]. One of the most highlighted topics in the FM/NM bilayer is a current-driven spin-orbit torque (SOT) [2,3,4], which is realization of charge to spin current conversion: the SOT is generated by the bulk spin Hall effect (SHE) [5] and/or interface Rashba-Edelstein effect (REE) [6], while both effects are triggered by the strong SOC. Of particular interest is the SOT via the SHE, which demonstrates magnetization switching of the adjacent FM [2,7,8]. For the last decade, the SHE-induced SOT has been further manifested to give rise to highly-efficient motion of chiral spin textures such as magnetic Néel domain walls [9–13] and skyrmions [14,15]. Many previous spintronic studies so far have mainly used 5$d$ transition metals (TMs) with the strong SOC as NM, such as Pt [2,9,10,12,13,16–18], Ta [7,9,11,14,19–23], and W [8,24,25].

5$d$ transition-metal oxides (TMOs) have recently emerged as a promising class of spintronic materials because of their unique electronic structure; the Fermi surface is dominated only by 5$d$ electrons with strong SOC, in contrast to that of 5$d$ TMs dominated by both 5$d$ and 6$s$ electrons. Efficient charge to spin-current conversion has been proven by a sizable SOT generation via the SHE in conductive 5$d$ TMOs such as Ir and W oxides: epitaxial $SrIrO_3$ [26–30], epitaxial $IrO_2$ [31,32], amorphous $IrO_2$ [33,34], and epitaxial $WO_2$ [35]. One of the advantages of 5$d$ TMOs is that their epitaxial form can be fully utilized, unlike most of the spintronics studies with 5$d$ TMs which are primarily focused on polycrystalline films. For instance, the SOT can be controlled in epitaxial Ir oxides by engineering the crystal orientation with the appropriate choice of substrates [28,31,32]. This contrast between epitaxial and polycrystalline forms raises an unsolved question how the crystallinity influences the SOT. Considering that the crystallinity is strongly linked to charge transport properties, e.g., electrical resistivity, the role of the crystallinity in the spin-transport properties is an intriguing subject. Engineering the crystallinity using the 5$d$ TMOs is therefore an effective approach to further understanding of the spin-current physics via the SOT generation. Among the 5$d$ TMOs, the particularly promising is the binary $IrO_2$, which can be a platform to realize the bilayer devices with different crystalline states such as amorphous, polycrystalline, and epitaxial forms. The epitaxial form, namely, rutile $IrO_2$ structure with a magnetic rutile oxide such as $CrO_2$, allows creation of a high-quality all oxide epitaxial interface, which is an excellent platform for clarifying intriguing spin-current physics; the $CrO_2$ is a room-temperature half metal [36,37].

In this paper, we report on the effect of the crystallinity on the SOT by fabricating the three states of the $IrO_2$. The SOT with dampinglike (DL) and fieldlike (FL) symmetries in $Co_{20}Fe_{60}B_{20}$ (CoFeB)/$IrO_2$ bilayers is characterized by harmonic Hall measurement. The DL SOT efficiency is enhanced with increase of the electrical resistivity, demonstrating that the spin-current properties are strongly affected by their crystallinity. Furthermore, the spin Hall conductivities of the three crystalline states of $IrO_2$ exhibit nearly constant values despite their different electrical conductivities, consistent with the intrinsic regime of the SHE scaling relation.

We fabricate the $IrO_2$ thin films by the reactive magnetron sputtering of an Ir target at a working pressure of 0.4 Pa with Ar and oxygen; the sputtering parameters are 75 W for the power and 20% for the partial oxygen pressure. For simplicity, we denote amorphous $IrO_2$, polycrystalline $IrO_2$, and epitaxial $IrO_2$ as a-$IrO_2$, poly-$IrO_2$, and epi-$IrO_2$. In order to selectively grow the three states, we chose a single-crystalline corundum $Al_2O_3$ (0001) substrate for epi-$IrO_2$ and thermally oxidized Si substrates for poly-$IrO_2$ and a-$IrO_2$; the thermally oxidized Si is prepared by coating a 500-nm-thick amorphous $SiO_2$ layer deposited on Si substrate. Fixed substrate temperature is at 350 ℃ for epi-$IrO_2$ and poly-$IrO_2$, and ambient temperature for a-$IrO_2$, respectively. We prepared the single layer $IrO_2$ films with its thickness of 30 – 40 nm for structural characterization by *x*-ray diffraction (XRD) and *x*-ray reflectivity (XRR) measurement. Figure 1(a) shows a schematic of crystal structure of rutile $IrO_2$ [38]. The lattice vector represents the rutile unit cell: $a = b > c$, where $a = b = 0.4498$ nm, and $c = 0.3154$ nm for bulk [38]. We draw a simplified sketch of surface oxygen–oxygen arrangement to aid in understanding the crystallographic relationship between $IrO_2$ and $Al_2O_3$ as discussed later. The $Al_2O_3$ has the hexagonal Bravais lattice with an in-plane lattice constant of 0.476 nm [39]. While the gray area represents the *a*-axis-oriented $IrO_2$ plane, the green, purple, and yellow areas correspond to its in-plane rotation of 30°, 150°, and 270° on the (0001)-oriented $Al_2O_3$ plane, respectively.

Figure 1(b) shows the out-of-plane XRD pattern of epi-$IrO_2$ grown on $Al_2O_3$(0001) substrate. We observe two peaks, (100) and (200), of the rutile $IrO_2$ without any impurity peaks. The inset of Fig. 1(b) shows the magnified view around the $IrO_2$(200) and the $Al_2O_3$(0006) peaks. The $IrO_2$(200) peak yields an out-of-plane lattice constant of 0.4623 nm, which is close to the previously reported value of ~0.4604 nm for the epitaxial $IrO_2$(100) grown on $TiO_2$(100) substrate [40]. The Laue fringes around the (200) diffraction peak evidence a good crystallinity of the film, which is also supported by the narrow full width at half maximum of the rocking curve around $IrO_2$(200) (~ 0.030 deg.). In order to understand the in-plane crystallographic orientation of the epi-$IrO_2$, we measured the azimuthal angle $\phi$ scan at the $IrO_2$(101) [top panel in Fig. 1(c)] and $Al_2O_3$($1\bar{2}12$) [bottom panel in Fig. 1(c)] diffractions. The three diffraction peaks of $Al_2O_3$($1\bar{2}12$) and its equivalents reflect the three-fold symmetry of the corundum structure [35,41]. The six-fold diffraction peaks of $IrO_2$(101) and its equivalent (110) were observed, which were rotated by 30° with respect to $Al_2O_3$($1\bar{2}12$). We therefore ascribe the observed six peaks of $IrO_2$ (101) and (110) to the formation of three crystalline domains with the epitaxial relationship of rutile $IrO_2[001](100)//Al_2O_3[10\bar{1}0](001)$ and the other two equivalent orientations as depicted in Fig. 1(a). Figure 1(d) shows the out-of-plane XRD patterns of poly-$IrO_2$ and a-$IrO_2$ grown on thermally oxidized Si substrates. Several diffraction peaks from $IrO_2$(110), (200), and (211) were observed for poly-$IrO_2$, while no significant peak was detected for a-$IrO_2$, confirming that crystallinity was able to be controlled by growth condition and substrate selection; we also confirmed that the valence states are both $Ir^{4+}$ for a- and poly-$IrO_2$, based on their mass densities obtained from the XRR measurements (see APPENDIX A). Moreover, poly-$IrO_2$ exhibits a crystallite size of ~ 7 nm, as determined from the full width at half maximum of the $IrO_2$(200) diffraction peak shown in Fig. 1(d) (see APPENDIX B). In contrast, the presence of Laue fringes in epi-$IrO_2$ suggests that its crystallite size is comparable to the film thickness (~30 nm). These results support distinct crystallinities between poly-$IrO_2$ and epi-$IrO_2$. Given this evidence of crystallinity, we thus demonstrated fabrication of $IrO_2$ films with three distinct crystalline states for investigating spin current generation. We prepared bilayers consisting of ferromagnetic $Co_{20}Fe_{60}B_{20}$ (referred as CoFeB hereafter) and $IrO_2$ to examine the effect of crystallinity of $IrO_2$ thin films on SOT. The whole film structures are represented by $TiO_x$(2) capped CoFeB(3)/$IrO_2$($t$) bilayer on substrate, where the number in parenthesis indicates the thickness in nanometer [Fig. 2(a)]. We vary the thickness of the $IrO_2$ layer $t$ from 4 to 10 nm to characterize its electrical transport properties. After the $IrO_2$ growth, Ti and CoFeB were deposited in-situ by a radio frequency magnetron sputtering at an Ar deposition pressure of 0.4 Pa. The $TiO_x$ layer was obtained by natural oxidation of the as-deposited Ti metal in the air; it prevents the CoFeB layer from oxidization. We confirmed no SOT contribution from the $TiO_x$ layer by control experiment. The CoFeB and $TiO_x$ thicknesses were estimated from the growth rate of each layer determined by the XRR measurement for thick reference samples. In comparison with the $IrO_2$, we also prepared an Ir control sample for proving the validity of our SOT measurement set-up, that is, $TiO_x$(2)/CoFeB(3)/Ir(6) on the thermally oxidized Si substrate. The CoFeB/$IrO_2$ and CoFeB/Ir bilayers were patterned into a Hall bar with two arms by photolithography and Ar ion milling. Ti(5)/Pt(60) contact pads were attached at the end of devices for electrical measurement. As shown in Fig. 2(a), channel length ($L$) and width ($w$) of the Hall-bar devices are $L = 30$ μm and $w = 10$ μm for epi-$IrO_2$, and $L = 50$ μm and $w = 10$ μm for the poly- and a-$IrO_2$ samples. We define $x$ and $y$ axis as longitudinal and transverse direction of the Hall bar devices. The $\phi_H$ represents the azimuthal angles of the external magnetic field ($B_{ext}$) in $xy$ plane. In order to examine anisotropic SOT [30, 31, 42] for epi-$IrO_2$, we patterned two orthogonal Hall bars along in-plane orientations [10$\bar{1}$0] and [1$\bar{2}$10] of the $Al_2O_3$ substrate, while the anisotropic effect can be averaged out by the multiple crystalline domains demonstrated in Fig. 1(c). We applied an ac current $I_{ac}=\sqrt{2}I_{rms}\sin(2\pi ft)$ with root mean square of current $I_{rms}$ and frequency $f = 13$ Hz. We set $I_{rms}$ to be 0.1 mA for longitudinal resistance ($R$) measurement and Hall resistance ($R_H$) measurement, and 1.5– 2.0 mA for harmonic Hall measurement.

We first characterized electrical transport properties of the IrO$_2$ layer by $t$ dependence of sheet conductance $L/(Rw)$ shown in Fig. 2(b); the linear fitting yields electrical conductivity $\sigma$ of the IrO$_2$ layer. We obtained the electrical resistivity $\rho\, (= 1/\sigma)$ of $\rho = 350$ μΩcm for a-IrO$_2$, 210 μΩcm for poly-IrO$_2$, and 140 μΩcm for epi-IrO$_2$. The $\rho$ values of epi-IrO$_2$ along the [10$\bar{1}$0] and [1$\bar{2}$10] of the Al$_2$O$_3$ substrate are equal, which is consistent with multi-domain nature of IrO$_2$ on Al$_2$O$_3$ as discussed in Fig. 1(c). Assuming the resistivity of CoFeB to be the previously obtained value of 165 μΩcm [30], we determined $\rho = 25$ μΩcm for Ir metal as control sample, which is comparable to previous reports [33,43]. Figure 2(c) summarizes the $\rho$ of the three states of IrO$_2$ layer, indicating that the $\rho$ monotonically decreases with the improvement in the crystallinity of IrO$_2$.

We independently quantify DL SOT and FL SOT by performing harmonic Hall measurements as described in Refs. [16,17,20]. By applying the $I_{ac}$ to the Hall bar, the SOT is induced at the FM/NM interface, which has two components with different symmetries, namely, DL and FL SOTs [3]. These SOTs correspond to DL effective field $B_{DL} \parallel (s \times m)$ and FL effective field $B_{FL} \parallel s$, where $m$ and $s$ represent the directions of the magnetization and accumulated spin polarization at the interface; the $s$ is oriented along the $y$-axis direction. Of these two SOTs, the DL SOT is only relevant to magnetization switching [2,3,4,7]. In order to distinguish the $B_{DL}$ and $B_{FL}$, the harmonic Hall measurements were performed by measuring the $\phi_H$ dependence of the first and second harmonic resistances ($R_H^{1\omega}$, $R_H^{2\omega}$) at $B_{ext}$ varying 0.087– 1.191 T. The $R_H^{1\omega}$ corresponds to the conventional dc Hall measurement, while the $R_H^{2\omega}$ reflects the influence of SOT; the former and the latter obey Eqs. (1) and (2) [16,17,20],

$$R_H^{1\omega} = R_{PHE} \sin 2\phi_H \quad (1)$$

$$R_H^{2\omega} = -\left(R_{AHE}\frac{B_{DL}}{B_{ext}+B_k}+R_{\nabla T}\right)\cos\phi_H + 2R_{PHE}\frac{B_{FL}+B_{Oe}}{B_{ext}}(2\cos^3\phi_H - \cos\phi_H) \quad (2)$$

$$\equiv -R_{DL+\nabla T}\cos\phi_H + R_{FL+Oe}(2\cos^3\phi_H - \cos\phi_H) \quad (3)$$

Here, $R_{PHE}$, $R_{AHE}$, $B_k$, $R_{\nabla T}$, and $B_{Oe}$ correspond to planar Hall resistance, anomalous Hall resistance, out-of-plane anisotropy field, thermal induced second-harmonic resistance, such as the anomalous Nernst effect [44] and spin Seebeck effect [45], and current induced Oersted field, respectively. $R_{AHE}$ and $B_k$ were estimated by out-of-plane magnetic field dependence of Hall resistance. The $\phi_H$ dependence of $R_H^{2\omega}$ is induced by the small modulation of the magnetization from its equilibrium position owing to the current-driven SOTs. We can define the DL contribution ($R_{DL+\nabla T}$) and the FL contribution ($R_{FL+Oe}$) as the coefficients of the $\cos\phi_H$ and $(2\cos^3\phi_H - \cos\phi_H)$ in Eq. (3). We now take a look on the experimental data. The upper panel of Fig. 3(a) shows the representative $R_H^{1\omega}$ as a function of $\phi_H$ in CoFeB(3)/epi- IrO$_2$(6) measured at 0.087 and 0.992 T, where the overall trend is reproduced by Eq. (1). A slight deviation of $R_H^{1\omega}$ from Eq. (1) is attributed to anomalous Hall effect (AHE) resulting from a small sample misalignment [20,28,30,34]. By confirming that the AHE contribution is within 3% on the $R_H^{1\omega}$, which is negligibly small compared with $R_{PHE}$, we thus determined the $R_{PHE}$ from fitting based on Eq. (1). The bottom panel of Fig. 3(a) shows the corresponding result of $R_H^{2\omega}$, which is well fitted by Eq. (2). The difference in amplitude of the $R_H^{2\omega}$ depending on the $B_{ext}$ indicates the suppression of SOTs by the $B_{ext}$. We estimated the $R_{DL+\nabla T}$ and $R_{FL+Oe}$ from the $R_H^{2\omega}$ fitting, which is displayed in the top panel and the bottom panel in Fig. 3(b), respectively. The result indicates the contribution of $R_{DL+\nabla T}$ is larger than that of $R_{FL+Oe}$ to the $R_H^{2\omega}$ in the CoFeB/epi-IrO$_2$ bilayer.

We extract two SOT effective fields $B_{DL}$ and $B_{FL}$ by analyzing magnetic field dependence of $R_{DL+\nabla T}$ and $R_{FL+Oe}$. Figure 3(c) shows the $R_{DL+\nabla T}$ as a function of $1/(B_{ext} + B_k)$ for CoFeB/epi-IrO$_2$, which obeys the first term in Eqs. (2) and (3). The slope and intercept of the $R_{DL+\nabla T}$ correspond to the $B_{DL}$ and the $R_{\nabla T}$, respectively. We determined the DL effective field per current density $B_{DL}/J$, where $J$ is the applied charge current density flowing in the IrO$_2$ layer. Based on the parallel resistor model as discussed in Fig. 2, we calculate $J$. The estimated $B_{DL}/J$ is +0.143 mT/(10$^{11}$Am$^{-2}$), suggesting the DL SOT generation in epi-IrO$_2$. The corresponding result of $R_{FL+Oe}$ is plotted as a function of $1/B_{ext}$ as shown in Fig. 3(d). The $R_{FL+Oe}$ has linear dependence on $1/B_{ext}$, which agrees with the second term in Eqs. (2) and (3). The $B_{FL}/J$ is determined to be $-0.126$ mT/(10$^{11}$Am$^{-2}$) by subtracting the $B_{Oe}$ contribution estimated by Ampere's law as $B_{Oe} = \mu_0 Jt/2$, where $\mu_0$ is the magnetic permeability in a vacuum.

We then evaluate the SOT efficiency $\xi_{DL(FL)}$ from $B_{DL(FL)}/J$ using the following equation [46]:

$$\xi_{DL(FL)} = \frac{2eM_s d}{\hbar} \frac{B_{DL(FL)}}{J} \tag{4}$$

where $e$, $M_s$, $d$, and $\hbar$ are the elementary charge, the saturation magnetization, the magnetic layer thickness, and the Dirac constant, respectively. The $M_s$ for all the films was obtained to be (0.92–1.0) × 10$^6$ A/m by using a superconducting quantum interference device magnetometer, in agreement with typical value of CoFeB thin films [7,8,11,19,24,25,43]. The $\xi_{DL}$ for epi-IrO$_2$ (100) with $J$//Al$_2$O$_3$ (10$\bar{1}$0) is determined to be +0.013. We performed the same experiment and analysis for the other samples. We estimate the $\xi_{DL}$ = +0.014 for epi-IrO$_2$ (100) with $J$ //Al$_2$O$_3$ (1$\bar{2}$10), +0.053 for poly-IrO$_2$, and +0.089 for a-IrO$_2$ as summarized in Fig. 3(e), while Ir metal shows +0.008 (not shown). The $\xi_{DL}$ values for all investigated IrO$_2$ and Ir films has different magnitude with the positive sign. The $\xi_{DL}$ for a-IrO$_2$ and Ir metal is in close agreement with the literature values [33,34,43], demonstrating the validity of our experimental set-up. Compared to other 5$d$ oxides, the $\xi_{DL}$ for IrO$_2$ (0.01–0.09) in the present study is lower than 0.15–0.55 for epitaxial SrIrO$_3$ [26,27,30] and +0.17 for epitaxial WO$_2$ [35]. Despite these relatively low $\xi_{DL}$ values, we have demonstrated the impact of the crystallinity of IrO$_2$ on SOT. The epi-IrO$_2$ (100) has the almost same magnitude of $\xi_{DL}$ for two current directions, indicating that the in-plane anisotropic SOT is smeared out by multi-domain formation. Judging from the negligible anisotropic effect of SOT and $\rho$, we hereafter consider the two devices as identical, resulting in the $\xi_{DL}$ for epi-IrO$_2$ determined as the average value of +0.014. In contrast to the epi-IrO$_2$, the magnitude of $\xi_{DL}$ for poly-IrO$_2$ consists of the averaged contributions from various crystal orientations [Fig. 1(d)]. The poly-IrO$_2$ is found to exhibit a smaller $\xi_{DL}$ compared to the a-IrO$_2$, giving the ratio of a-IrO$_2$/poly-IrO$_2$ ~ 1.68. Previous study has shown the corresponding ratio of ~1.63 in the spin Hall angle $\theta_{SH}$, with 0.065 for a-IrO$_2$ and 0.040 for poly-IrO$_2$, extracted from the inverse SHE technique based on non-local transport measurement [47]. Given the conventional formula $\xi_{DL} = T_{int}\theta_{SH}$, the observed coincidence between $\xi_{DL}$ and $\theta_{SH}$ reveals that the spin transparency $T_{int}$ at the interface of IrO$_2$ and CoFeB is nearly the same for a-IrO$_2$ and poly-IrO$_2$. Figure 3(f) shows the $\xi_{DL}$ as a function of the $\rho$ for the epi-, poly, and a- IrO$_2$ layers. The $\xi_{DL}$ and $\rho$ concomitantly increase in order of epi-IrO$_2$, poly- IrO$_2$, and a- IrO$_2$, indicating the significant role of crystallinity on the spin-current and electrical transport properties. We will further discuss the trend between the $\xi_{DL}$ and the $\rho$ later. Regarding the FL SOT, we obtain $\xi_{FL}$ = −0.010 for the epi-IrO$_2$ (100), −0.022 for poly-IrO$_2$, −0.015 for a-IrO$_2$, and +0.004 for Ir metal. The $\xi_{FL}$ in all the IrO$_2$ samples are negative and their magnitudes are smaller than those of the corresponding $\xi_{DL}$, which exhibit positive signs; this difference in sign suggests that the origin of FL SOT is distinct from that of the DL SOT. While it is sometimes attributed to interfacial REE in metallic bilayers [16,17,19,20,24], its precise origin remains under active debate in spintronics. The larger $\xi_{DL}$ than $\xi_{FL}$ is helpful for demonstrating the magnetization switching since the DL SOT plays an important role in magnetization control.

We now discuss the SHE scaling relation to gain further insight into how $\xi_{DL}$ correlates with the crystallinity. The SHE scaling relation has been much less experimentally investigated than the AHE mainly because the SHE is achieved only in metals with high conductivity. It is theoretically proposed that the SHE obeys the same scaling relation as the AHE since both the SHE and the AHE originate from the intrinsic and the extrinsic mechanisms arising from the SOC [48,49]. In the widely investigated AHE scaling relation between the anomalous Hall conductivity $\sigma_{AH}$ and the electrical conductivity $\sigma$ (= $1/\rho$), the AHE is classified into three regimes [50]: dirty ($\sigma_{AH} \propto \sigma^{1.6}$), intermediate (intrinsic, $\sigma_{AH}$ = constant), and clean (extrinsic, $\sigma_{AH} \propto \sigma$). Based on the concept, we plot the spin Hall conductivity $\sigma_{SH}$ as a function of $\sigma$ for the three crystalline forms of IrO$_2$ in Fig. 4, with the AHE scaling relation indicated by black line [50] and with the $\sigma_{SH}$ for polycrystalline Pt deposited at various growth rate [51]. Here, the $\sigma_{SH}$ for IrO$_2$ is defined by $\sigma\xi_{DL}$ with the assumption of $T_{int} = 1$, which is similar to previous reports [15,23,31]. The $\sigma_{SH}$ for a-IrO$_2$ and poly-IrO$_2$ coincides at $2.5 \times 10^2$ $\Omega^{-1}$cm$^{-1}$, whereas that for epi-IrO$_2$ ($1.0 \times 10^2$ $\Omega^{-1}$cm$^{-1}$) is smaller than them. Since the theoretical scaling relationship neglects anisotropy of materials [50], we have to consider average of anisotropic spin Hall conductivities when comparing our results with the theory; we judged that the poly-IrO$_2$ takes the average value of $\sigma_{SH}$ because it consists of several crystal orientations discussed in Fig. 1(d). In contrast, previous studies have demonstrated the growth of epitaxial IrO$_2$ films on lattice-matched TiO$_2$ substrates with various crystal orientations, revealing that (001)-oriented IrO$_2$ exhibits twice larger spin Hall conductivity than (110)-oriented IrO$_2$ [31,32]. Considering that IrO$_2$ has a tetragonal rutile structure with $c$-axis as a fourfold rotational screw axis, we expect that $\sigma_{SH}$ of our (100)-oriented epi-IrO$_2$ is close to that of (110)-oriented IrO$_2$ and hence is smaller than that of (001)-oriented IrO$_2$; hypothetical averaged $\sigma_{SH}$ for epi-IrO$_2$ might be larger than the observed value of $1.0 \times 10^2$ $\Omega^{-1}$cm$^{-1}$. In addition to the spin Hall conductivity, the reported crystal-orientation dependence of electrical conductivity in IrO$_2$ films grown on TiO$_2$ substrates, ranging from $3.85 \times 10^3$ to $1.25 \times 10^4$ $\Omega^{-1}$cm$^{-1}$, indicates a variation by a factor of approximately 3 [40]. This also suggests that the hypothetical averaged conductivity expected for single-crystalline IrO$_2$ is unlikely to be significantly higher than that of our film. These findings indicate that the anisotropies in both spin Hall conductivity and electrical conductivity are not large enough to cause a substantial deviation in the log-log plot in Fig. 4. Given the discussion, although our epitaxial IrO$_2$ film is not perfectly single-crystalline in the in-plane direction, our experimental approach remains valid, since performing SOT and conductivity measurements for all possible crystal orientations would not yield substantially more insight beyond what is already discussed. We can therefore conclude that the three data points for IrO$_2$ are nearly constant, which is consistent with the intrinsic regime. The $\sigma_{SH}$ for the Pt is almost constant for their wide range of $\sigma$, again indicating the intermediate regime. In our results for IrO$_2$, we can extend the $\sigma$ down to $3 \times 10^3$ $\Omega^{-1}$cm$^{-1}$, which is close to the boundary between the dirty and intermediate regimes. Our results experimentally reinforce the agreement between the SHE and the AHE scaling relationship, even in the less conductive compounds such as amorphous IrO$_2$.

In conclusion, we studied the SOT generation in the CFB/IrO$_2$ bilayers, where the three crystalline states of the IrO$_2$ layer were prepared by the reactive sputtering method. The structures of the epi-IrO$_2$ (100), poly-IrO$_2$, and a-IrO$_2$ layers were characterized by the XRD. By conducting harmonic Hall measurement, we estimated the $\xi_{DL}$ values of +0.014 for epi-IrO$_2$, +0.053 for poly-IrO$_2$, and +0.089 for a-IrO$_2$, which monotonically increase with their respective $\rho$ values. These results indicate that engineering crystallinity provides an efficient route for controlling the SOT. Moreover, the $\sigma_{SH}$ remains nearly constant with respect to their $\sigma$, highlighting that the intrinsic regime of the SHE scaling relation is dominant in the spin-current property. This work paves a new pathway for engineering the charge to spin-current conversion efficiency in 5$d$ TMOs, leading to the development of low-power spintronic devices.

## APPENDIX A: Discussion for valence state of Ir$^{4+}$ in amorphous IrO$_2$ and polycrystalline IrO$_2$ films using mass density measurements obtained through XRR

As shown in Fig. 5(a), the wide-range XRR profiles for amorphous and polycrystalline IrO$_2$ films grown on Si substrates exhibit similar overall reflectivity features, indicating comparable thicknesses and density ranges. To quantitatively extract the film parameters, we fitted the experimental XRR data in Fig. 5(b) using a theoretical model in which a uniform, single-composition IrO$_2$ film is grown on a SiO$_2$ substrate with zero surface roughness. The fitting was performed using the GenX software package [52]. Among the fitting parameters such as film surface roughness, film thickness, and mass density, the primary focus was on determining the mass density with high accuracy. The best-fit results yielded a thickness of approximately 35 nm and a surface roughness of 0.5 nm for both films. We obtained the fitted mass densities for $\rho_M$ (a-IrO$_2$) = 11.0 g/cm³ and $\rho_M$ (poly-IrO$_2$) = 11.9 g/cm³, indicating that the amorphous film is ~8% less dense than the polycrystalline one. This finding is consistent with theoretical predictions; although the unit-cell volume of a-IrO$_2$ cannot be easily measured, a prior theoretical report suggests a 9% lower mass density for amorphous IrO$_2$ (10.7 g/cm³) compared to crystalline IrO$_2$ (11.7 g/cm³) [53]. The low-angle XRR profiles in Fig. 5(c) further emphasize this difference, with red and blue triangles indicating the critical angles for a-IrO$_2$ and poly-IrO$_2$, respectively. These critical angles were calculated from the fitted mass densities using the following relation:

$$\theta_c = \sqrt{\frac{r_e \lambda^2}{\pi} \cdot \frac{Z}{A} \cdot N_A \cdot \rho_M}$$

where $\theta_c$ is the critical angle, $r_e$ is the classical electron radius, $\lambda$ is the x-ray wavelength, $Z$ is the total number of electrons per formula unit, $A$ is the molar mass of the formula unit, $N_A$ is Avogadro's number, and $\rho_M$ is the mass density. The good agreement between the calculated critical angles and experimental data further supports the reliability of the fitting results. Given the agreement between the fitted densities and theoretical expectations, and assuming the same IrO$_2$ stoichiometry, we conclude that the oxygen content is equivalent within experimental error. This supports the view that both films predominantly contain valence state of Ir$^{4+}$.

## APPENDIX B: Distinct crystallinities between polycrystalline and epitaxial IrO$_2$ films

To provide further insight into crystallinity, we estimated a crystallite size of ($D$) for poly-$IrO_2$ according to Scherrer equation via $D = 0.9\lambda/(w\cos\theta)$, where $\lambda$ is the x ray wave length, $w$ is the full width at half maximum of diffraction pattern, and is the Bragg's angle [54]. Figure 6 represents the magnified view of x ray diffraction pattern for poly-$IrO_2$(200) film. By determining $w = 1.23$ from black fitting curve, we estimated $D \sim 7$ nm, which is significantly smaller than total thickness of 30 nm. In contrast, the presence of Laue fringes in epi-$IrO_2$ in Fig. 1(b) suggests that its crystallite size is comparable to the film thickness (~30 nm). These results provide supportive information for distinct crystallinity between poly- and epi-$IrO_2$ films.

## ACKNOWLEDGEMENT

The authors thank T. Arakawa for technical support. This work was partly carried out at the Center for Advanced High Magnetic Field Science in Osaka University under the Visiting Researcher's Program of the Institute for Solid State Physics, the University of Tokyo. This work was partly supported by Nanotechnology Platform of MEXT, Grant Number JPMXP09S21OS0027, the JSPS KAKENHI (Grant Nos. JP19K15434, JP19H05823, 22H04478, and 24H01666), JPMJCR1901 (JST-CREST), the MEXT "Spintronics Research Network of Japan (Spin-RNJ)", Nippon Sheet Glass foundation for Materials Science and Engineering, and Iketani Science and Technology Foundation. We acknowledge stimulating discussions at the meeting of the Cooperative Research Project of the Research Institute of Electrical Communication, Tohoku University.

**ORCID iDs**
Kohei Ueda https://orcid.org/0000-0003-4967-2249
Junichi Shiogai https://orcid.org/0000-0002-5464-9839
Takanori Kida https://orcid.org/0000-0003-3650-3534
Masayuki Hagiwara https://orcid.org/0000-0002-1087-521X
Jobu Matsuno https://orcid.org/0000-0002-3018-8548

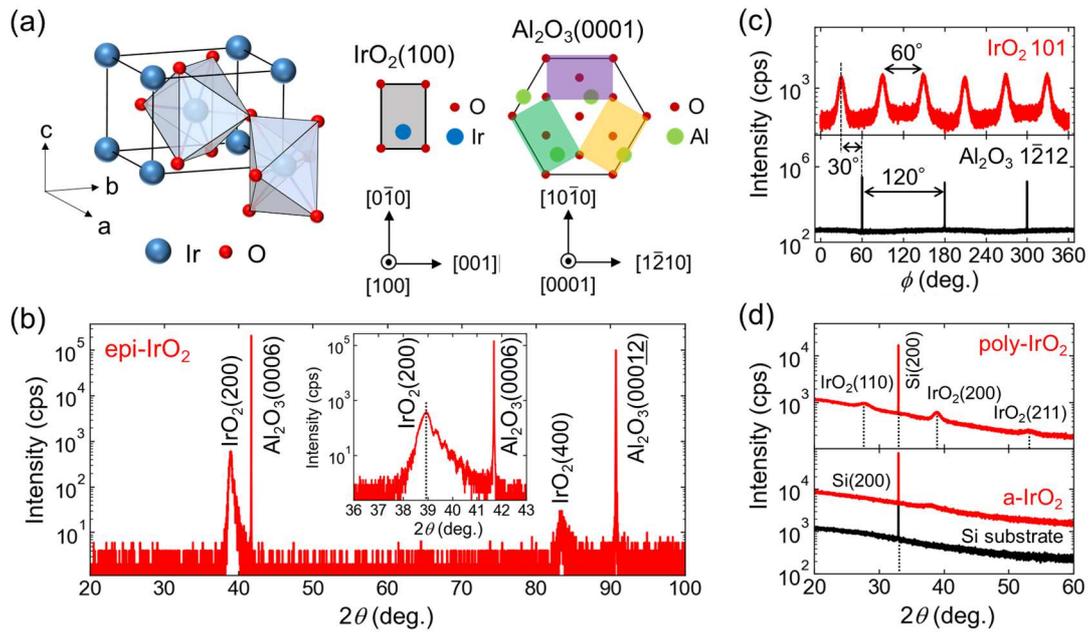

**Fig. 1.** (a) (left) Illustration of structure for rutile $IrO_2$. (right) The crystallographic relationship between the rutile-type $IrO_2$(100) and the $Al_2O_3$(0001). The purple, green, and yellow indicate three-type $IrO_2$ domains stemming from the trigonal symmetry of the $Al_2O_3$ substrate. (b) X-ray diffraction (XRD) $2\theta$-$\theta$ scan of epitaxial $IrO_2$ film grown on $Al_2O_3$ (0001) substrate. Magnified view of the XRD scan around epitaxial $IrO_2$(200) reflection grown on the $Al_2O_3$(0001) substrate (Inset). (c) The azimuthal $\phi$ scan of ($1\bar{2}12$) diffraction for the $Al_2O_3$ substrate (bottom) and of (101) diffraction for $IrO_2$ film (top). (d) The XRD pattern of polycrystalline $IrO_2$ (top), and amorphous $IrO_2$ and thermally oxidized Si substrate (bottom).

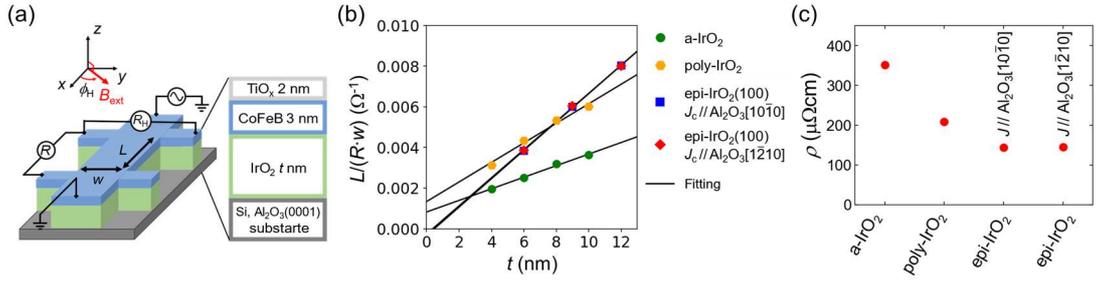

**Fig. 2.** (a) Schematic illustration of the device structure and layer geometry of the heterostructure samples. The $\phi_H$ represents the azimuthal angles of the external magnetic field ($B_{ext}$). AC current ($J$) is applied along the $x$ axis direction. Hall resistance ($R_H$) and longitudinal resistance ($R$) are measured along $y$-axis and $x$-axis direction, respectively. (b) Sheet conductance as a function of the $IrO_2$ thickness ($t$) for amorphous, polycrystalline, and epitaxial $IrO_2$ with 3-nm-thick CoFeB layer. The solid lines show linear fitting on the data. (c) Electrical resistivity for amorphous, polycrystalline, and epitaxial $IrO_2$. The direction of the $J$ for epitaxial $IrO_2$ is along $[10\bar{1}0]$ and $[1\bar{2}10]$ orientations on the $Al_2O_3$ substrate.

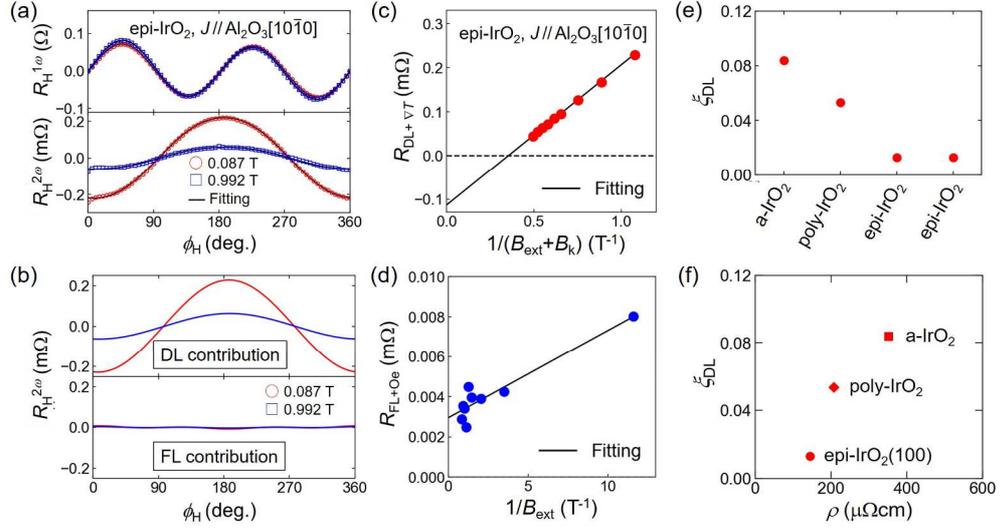

**Fig. 3.** (a) First-harmonic Hall resistance ($R_H^{1\omega}$) of 3-nm-thick CoFeB/6-nm-thick IrO$_2$ measured at 0.087 and 0.992 T with $J$ along Al$_2$O$_3$($10\bar{1}0$) (top). The solid curve is fit to the data using Eq. (1). The corresponding second-harmonic Hall resistance ($R_H^{2\omega}$) with solid curve being fit using Eq. (2) (bottom). (b) DL contribution ($\cos\phi_H$) and FL contribution ($2\cos^3\phi_H - \cos\phi_H$) on the $R_H^{2\omega}$, separated from the fitting result. (c) The $R_{DL+\nabla T}$ as a function of $1/(B_{ext} + B_k)$ and (d) the $R_{FL+Oe}$ as a function of $1/B_{ext}$. The solid line is linear fit to the experimental data. (e) DL SOT efficiency for amorphous, polycrystalline, and epitaxial IrO$_2$. The direction of the $J$ for epitaxial IrO$_2$ is along $[10\bar{1}0]$ and $[1\bar{2}10]$ orientations of the Al$_2$O$_3$ substrate. (f) Relation of $\xi_{DL}$ between $\rho$ for epi-, poly-, and a-IrO$_2$.

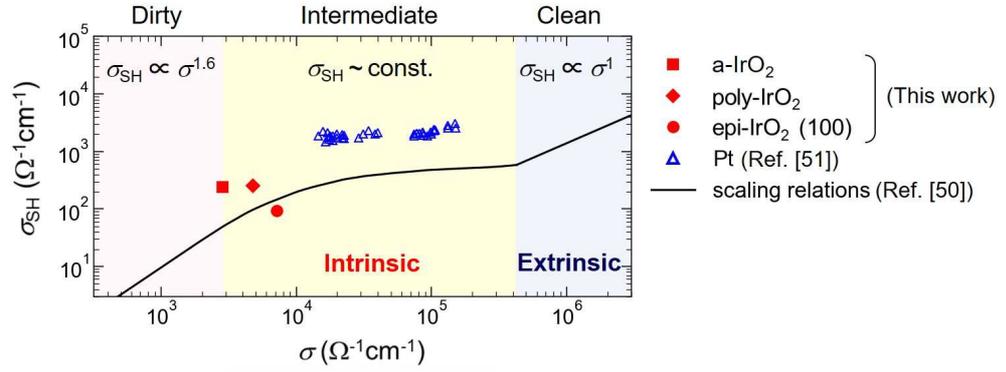

**Fig. 4.** The spin Hall effect scaling relation between spin Hall conductivity versus electrical conductivity in amorphous, polycrystalline, and epitaxial states of IrO$_2$, in comparison with polycrystalline Pt metal [51]. The black line represents the anomalous Hall effect scaling relation [50], which is dominated by the dirty, intermediate, and clean regimes.

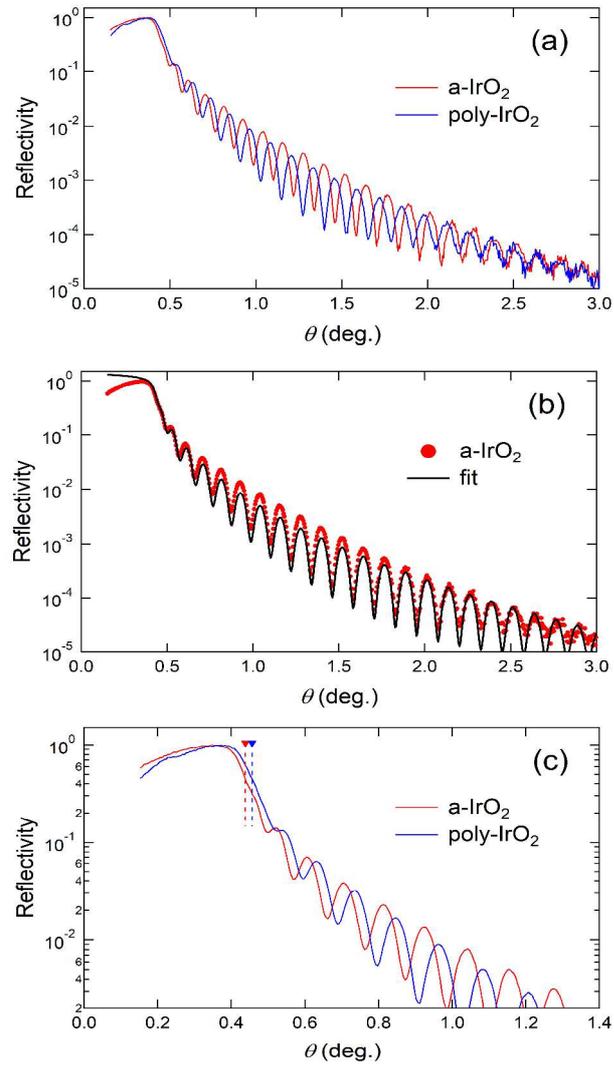

**Fig. 5.** (a) Wide-range x-ray reflectivity (XRR) profiles of $IrO_2$ films in amorphous (a-$IrO_2$) and polycrystalline (poly-$IrO_2$) states. (b) XRR profile of a-$IrO_2$ with the fitted curve. (c) Low-angle XRR profiles of amorphous and polycrystalline $IrO_2$, with critical angles indicated by red and blue triangles for a-$IrO_2$ and poly-$IrO_2$, respectively.

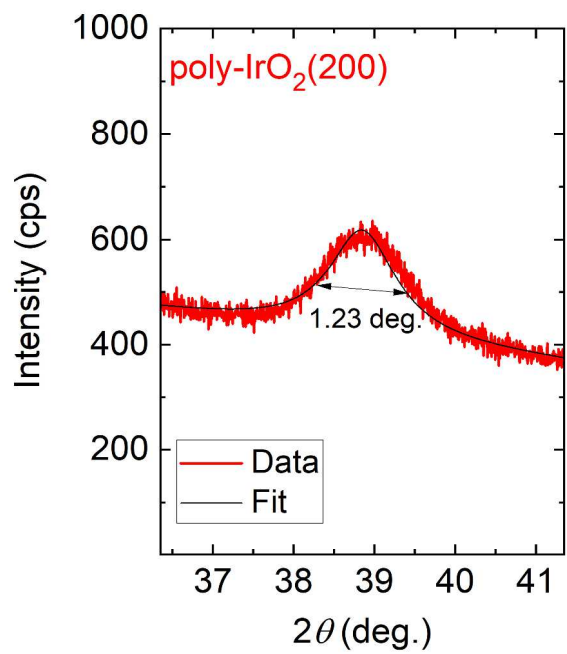

**Fig. 6.** Magnified view of x ray diffraction pattern for poly-IrO$_2$(200). Black solid curve represents a fit on data.